\documentclass{article}
\usepackage{spconf}
\usepackage{amsmath}
\usepackage{graphicx}
\usepackage{todonotes}
\usepackage{amssymb}
\usepackage{enumitem}
\usepackage{multirow}
\usepackage{siunitx}
\usepackage[ruled,vlined]{algorithm2e}
\usepackage{setspace}
\usepackage{hyperref}
\title{Ultra-Lightweight Neural Differential DSP Vocoder for High Quality Speech Synthesis}
\name{Prabhav Agrawal$^{*}$, Thilo Koehler$^{*}$\thanks{\text{}$^{*}$ means equal contribution, Correspondence to \{prabhavag, tkoehler, qinghe\}@meta.com}, Zhiping Xiu, Prashant Serai, Qing He}
\address{Meta AI}

\begin{document}
\ninept
\maketitle

\begin{abstract}
Neural vocoders model the raw audio waveform and synthesize high-quality audio, but even the highly efficient ones, like MB-MelGAN and LPCNet, fail to run real-time on a low-end device like a smartglass. A pure digital signal processing (DSP) based vocoder can be implemented via lightweight fast Fourier transforms (FFT), and therefore, is a magnitude faster than any neural vocoder.  A DSP vocoder often gets a lower audio quality due to consuming over-smoothed acoustic model predictions of approximate representations for the vocal tract.  In this paper, we propose an ultra-lightweight differential DSP (DDSP) vocoder that uses a jointly optimized acoustic model with a DSP vocoder, and learns without an extracted spectral feature for the vocal tract. The model achieves audio quality comparable to neural vocoders with a high average MOS of 4.36 while being efficient as a DSP vocoder. Our C++ implementation, without any hardware-specific optimization, is at 15 MFLOPS, surpasses MB-MelGAN by 340 times in terms of FLOPS, and achieves a vocoder-only RTF of 0.003 and overall RTF of 0.044 while running single-threaded on a 2GHz Intel Xeon CPU. 

\end{abstract}

\section{Introduction}
Synthesizing artificial speech from text, referred to as speech synthesis or text-to-speech (TTS), is the primary interface for AI voice assistants, in-car navigation systems, and accessibility devices for the visually impaired. With the increasing popularity of wearables such as smartwatches and smartglasses, there has been more demand for having on-device TTS to support private use cases of calling and messaging.

In the past few years, auto-regressive neural vocoders such as WaveNet \cite{oord2016wavenet}, SampleRNN \cite{mehri2017samplernn}, and WaveRNN \cite{kalchbrenner2018efficient} had tremendous success in generating high-fidelity realistic human voices, with no significant audio quality gap to actual recordings in subjective evaluations. They are suitable for server-based environments but are not efficient for on-device TTS due to model size and computational requirement of several GFLOPS. Parallel frame-wise audio sample generation in non-auto-regressive neural vocoders such as MelGAN \cite{kumar2019melgan}, HiFi-GAN \cite{kong2020hifi}, and WaveGlow \cite{prenger2018waveglow} can achieve higher synthesis speed than the auto-regressive sample-wise prediction, albeit only on GPU or multi-core CPU devices, since they do not reduce the model size or absolute computational complexity. Smaller state-of-the-art neural vocoders such as MB-MelGAN \cite{yang2021multi} or LPCNet \cite{valin2019lpcnet} start at around 3 GFLOPS, which are feasible for high-end mobile devices but far from ideal for battery life and memory usage for low-end wearable devices.

Neural vocoders that directly model the audio waveform are computationally intensive because modeling the phase of a waveform is challenging due to its stochastic nature. As described here \cite{engel2020ddsp}, different phase waveforms can sound the same, whereas waveforms with different magnitude spectrograms sound different. This observation motivates us to learn only the magnitude spectrograms well by comparing them against that of the true audio while procedurally generating phase information for efficiency.

In this paper, we propose a novel DDSP vocoder where we combine a simple and efficient DSP vocoder with the  acoustic model, described in Section \ref{sssec:dsp_vocoder}. The acoustic model is a neural net while DSP vocoder does not have any learnable parameters. Since the joint module is end-to-end differentiable, it can learn from the magnitude spectrogram of true audio.  Our DDSP vocoder achieves audio quality comparable to state-of-the-art neural vocoders, with the vocoder having a compute of 15 MFLOPS and vocoder-only RTF of 0.003 running single-threaded on a 2GHz Intel Xeon CPU.

Related DDSP works, such as neural homomorphic vocoder (NHV) \cite{Liu2020NeuralHV}, use a separate model to predict log-Mel spectrograms and then use a neural network to convert it to linear time-varying filter coefficients of the spectral envelope. NHV does an explicit modeling of phase for the spectral filters, whereas in this paper, we show that we can achieve high audio quality (4.36 avg MOS) with only zero-phase filters, with a reduction in vocoder FLOPS by 24 times over the NHV work.  Based on our knowledge, this is the first work where we see an acoustic model jointly trained end-to-end with a simple DSP vocoder with no learnable parameters using differential DSP techniques for optimization.

\begin{figure}
\vspace{-0.1in}
\centerline{\includegraphics[width=\columnwidth]{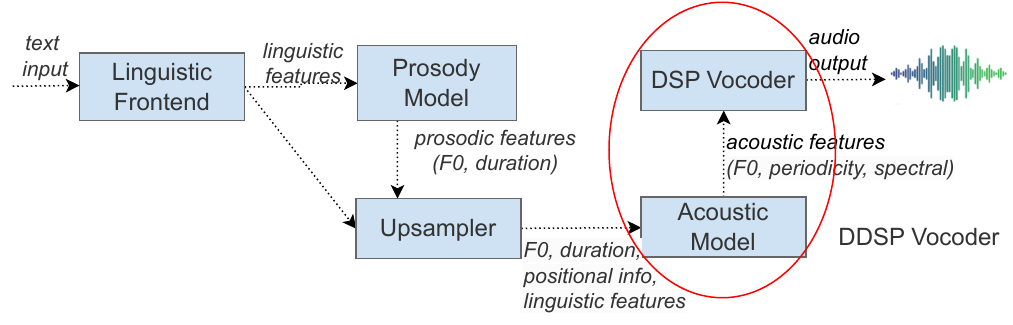}}
\caption{Our on-device TTS pipeline: The frontend extracts linguistic features, and the prosody model consumes them to output phone-level F0 and duration. Subsequently, the acoustic model takes the upsampled features,  to predict frame-level acoustic features, converted to audio waveform by DSP vocoder. In the DDSP vocoder, the acoustic model and DSP vocoder are combined into one single module.}
\label{fig:overview}
\vspace{-0.2in}
\end{figure}
\section{Proposed On-device TTS System}
\label{sec:tts_components}
In this section, we first describe the frontend components of on-device text-to-speech pipeline as shown in Figure \ref{fig:overview}, and then later describe the DDSP vocoder.

\subsection{Frontend Components}
\textbf{Linguistic Frontend}: Responsible for converting input text into linguistic features. It first normalizes the text, predicts the phonetic transcription using the International Phonetic Alphabet (IPA) \cite{kiel2015ipa} and then converts phones, syllable stress, and other supra-segmental information such as phrase type into one-hot features \cite{he2021}. It also adds pre-trained word embeddings \cite{ammar2016massively} \cite{upadhyay-etal-2016-cross} to improve the naturalness of the prosody. Features at phrase, word, or syllable rate are repeated for each phone to obtain one feature vector per phone.\\
\noindent
\textbf{Prosody Model}: Takes the linguistic features provided by the frontend and predicts the duration and the average fundamental frequency of each phone. The network architecture is an emformer\cite{shi2021}, with a  linear layer at input and two linear layers before the output, trained with an L2 loss on reference features estimated on the ground truth audio.\\
\noindent
\textbf{Upsampler}: Uses the phone-wise duration information to roll out the linguistic features into time synchronous frames by repeating them. It also includes the pitch and duration values, along with the positional information of the current frame within the current phone, syllable, word, and phrase.

\subsection{DDSP Vocoder}
\label{sec:ddsp_voc}
The DDSP vocoder consists of an acoustic model and a differential DSP vocoder, which are trained end-to-end with losses on the final audio waveform. In this section, we first describe the DSP vocoder and the acoustic model architectures separately. We then explain the end-to-end joint training procedure.

\subsubsection{DSP Vocoder}
\label{sssec:dsp_vocoder}
 \begin{figure}
\centerline{\includegraphics[width=0.4\textwidth]{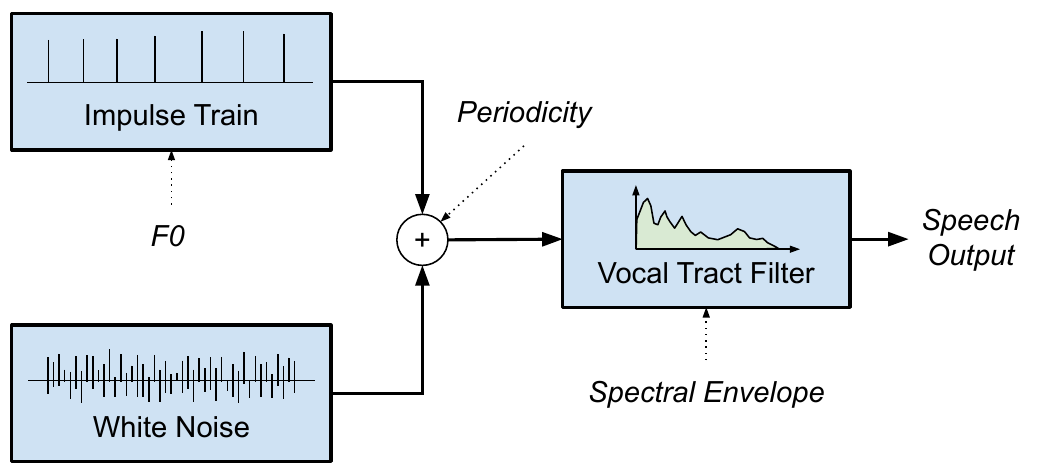}}
\caption{Source-Filter Model: The speech signal is generated by mixing an impulse train and white noise according to periodicity, followed by a filter representing the vocal tract and lip radiation.}
\label{fig:source_filter_model}
\vspace{-0.2in}
\end{figure}
Our DSP vocoder is based on the source-filter model \cite{rabiner2010theory} for speech production, as shown in Figure \ref{fig:source_filter_model} and takes three input features to generate the output speech signal $s$:
\begin{enumerate}
\vspace{-0.1cm}
\item Fundamental Frequency $F0$ (1-dim Hz value)
\vspace{-0.1cm}
\item Periodicity $P$ (12-dim mel band-wise ratio between periodic (only impulse train) and aperiodic excitation (only noise)) \cite{yoshimura2001mixed}
\vspace{-0.1cm}
\item Vocal Tract Filter $V$ (257-dim linear frequency log magnitude) 
\end{enumerate}
For the excitation signal $E$, it is either an impulse train $E_{imp}(F0)$ or white noise $E_{noise}$ of the same energy \cite{morise2016world} \cite{griffin1985new} \cite{KAWAHARA1999187}. Instead of combining them to get the mixed excitation signal, we split the vocal tract filter into the periodic and aperiodic parts by multiplying them with the periodicity feature. We then filter both excitation signals with their filters and add them to the final audio $s$. This approach allows us to optimize the algorithm used for each excitation type to avoid artifacts and make it computationally efficient. The equations describe the approach with uppercase denoting the variable in frequency domain vs lowercase denoting it in time domain.
\begin{equation}
    \label{eq:vocoder}
    \begin{alignedat}{3}
        s &= iFFT(E \times V) \\
        s &= iFFT([P \times E_{imp}(F0)  +  (1-P) \times E_{noise}] \times V) \\
        s &= \underbrace{iFFT(P \times V) \ast e_{imp}(F0)}_{\text{Periodic signal}}  +  \underbrace{iFFT((1-P) \times V \times E_{noise})}_{\text{Aperiodic signal}}
    \end{alignedat}
\end{equation}
Our vocoder generates audio at a sample rate of 24000 Hz that is merged via overlap-and-add to get the final audio waveform \cite{rabiner2010theory}. We choose a frame shift of 128 samples and an FFT size of 512 points. 12-dim $P$ is extrapolated to 257 linear coefficients. With 512 points, we can model frequencies down to $24000 Hz / 512 \approx 47 Hz$, which is sufficient for human speech. We allocate a 512+128 sample buffer for the periodic signal and a 512 sample buffer for the aperiodic signal. The periodic and aperiodic signals for each frame $i$ are then separately generated as follows:

\textbf{Periodic Signal:} We multiply the periodicity $P_i$ with the vocal tract filter $V_i$ to get the periodic part $P_i \times V_i$. Then, we convert $P_i \times V_i$ to the time domain using the inverse FFT and a phase of \SI{180}{\degree}. This represents a single impulse filtered by the periodic part of the filter $P_i \times V_i$. Then, we render the filtered impulse train by calculating the time stamps of the impulse within the frame by incrementing a running phase value by $1/F0_i$. It is multiplied by $1/sqrt(F0_i)$ for energy normalization. Note that it is possible that no impulse falls within the frame at low $F0_i$ values or the periodicity is entirely 0. In that case, we can skip the frame.

\textbf{Aperiodic Signal:} We shift the noise buffer by frame shift of 128 and fill the new 128 values with uniformly distributed pseudo-random numbers between $-1...+1$, multiplied with $1/sqrt(24000)$ to scale the noise to the same energy level as the impulses. We then convert the noise buffer to the frequency domain using forward FFT without any windowing function to get the complex spectrum $E_{noise_i}$. Note that windowing is not required for $E_{noise_i}$, since each sample is uncorrelated, which makes $E_{noise_i}$ perfectly periodic without any discontinuities. We multiply $E_{noise_i}$ with the aperiodic part of the filter $V_i \times (1-P_i)$. The result is then converted back to the time domain using inverse FFT. We then apply a centered Hann window \cite{rabiner2010theory} of size 256 points to intermediate noise buffer, so that overlapped audio can sum upto 1.0.

For each frame, both the intermediate audio buffers are then overlapped and added, with a frame shift of 128 samples to the final audio waveform.

\subsubsection{Acoustic Model}
\label{sssec:am}
The acoustic model consumes a 512-dim input vector of linguistic features, repeated phone-level F0 and duration, and the positional information for each frame. It follows an emformer architecture\cite{shi2021}; see Table \ref{tab:am_arch}, and gives a 270-dim output, which corresponds to 1-dim $F0$, 12-dim periodicity $P$ and a 257-dim representation for the vocal tract $V$. 

\begin{table}[!]
 \vspace{-0.1in}
    \centering
    \caption{Acoustic model architecture}
    \label{tab:am_arch}
    \resizebox{0.47\textwidth}{!}{
        \begin{tabular}{c | c}
            \hline
            Layer & Details \\
            \hline
            Linear + Tanh + Dropout(0.1)& in\_dim=512, out\_dim=128 \\
            \multirow{3}{*}{4 x Emformer} & in\_dim=128, out\_dim=128, ffn\_dim=512, \\
            & memory=4, seg\_size=32, left\_context=12,\\
            &  right\_context=12\\
            Linear + Tanh + Dropout(0.1) & in\_dim=128, out\_dim=199\\
            Linear & in\_dim=199, out\_dim=270 \\
            \hline
        \end{tabular}
    }
    \vspace{-0.2in}
\end{table}

\subsubsection{Joint Modeling via DDSP}
\label{ssec:diffvocmot}
\begin{figure}
\centerline{\includegraphics[width=0.50\textwidth]{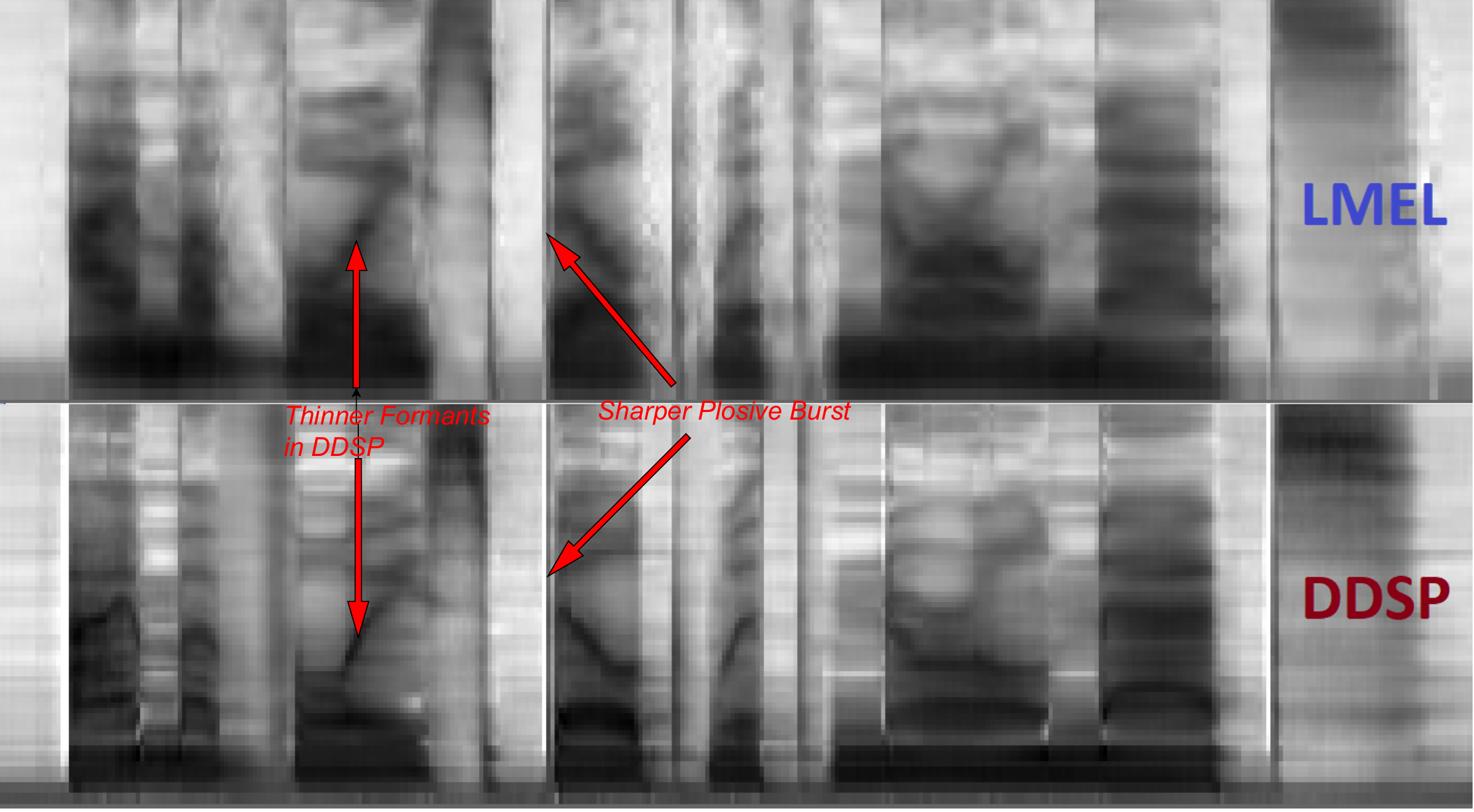}}
\caption{80-dim $lmel_{psync}$ prediction of our DSP Vocoder Adv (section \ref{ssec:expsetup}) vs learned spectral feature from DDSP vocoder with sharper formants and plosives in inverse grey coloring}
\label{fig:lmel_vs_ddsp}

\vspace{-0.5cm}
\end{figure}

As discussed in Section \ref{sssec:dsp_vocoder}, the excitation signal $E$ contains the phase information, while $V$ is a linear filter on top of E. Since we only observe the speech signal $s$, we can't accurately determine V, without knowing $E$. Methods to determine $V$ in literature,  based on cepstral smoothing, linear predictive coding (LPC) extraction or pitch synchronously extracted log mel spectrograms ($lmel_{psync}$), assume that V is responsible for slow changes throughout the magnitude spectrogram (formants), and create a smoothed version of the magnitude spectrogram of $s$ \cite{diemo1999spectral}\cite{huang2001spoken}. 
When we train the acoustic model to $lmel_{psync}$ features, the prediction errors also add up on top of approximate feature extraction. As a result, the audio sounds muffled and unnatural, and is penalized in subjective evaluations.  Since the DSP vocoder is differentiable, we can combine it together with the acoustic model. The setup can be jointly optimized by comparing the predicted audio against the true audio. This ensures that the spectral feature driving the vocoder is learned instead of engineered, and is optimized via true audio. 
Figure \ref{fig:lmel_vs_ddsp} shows a comparison of two acoustic model outputs, one is the  $lmel_{psync}$ prediction from DSP Vocoder Adv (see Section \ref{ssec:expsetup}, and the intermediate spectral representation learnt from DDSP vocoder, converted to 80-dim $lmel$ for comparison. We can see that DDSP vocoder has learnt a detailed spectral representation with thinner formants and sharper plosives.

\vspace{-0.2cm}

\subsubsection{Training }
\label{sssec:diffvocloss}
We used three types of losses to train our DDSP vocoder. Window size is kept the same as the FFT size for all audio feature extractions and loss calculations.

\textbf{Reference MSE Loss (on acoustic model predictions):} To get the training convergence, we apply an L2 loss for fundamental frequency prediction $\tilde{F0}$ with reference $F0$. For periodicity feature prediction $\tilde{P}$, we found that the system could learn it without explicit supervision from the reference $P$; however, having an L2 loss with the reference $P$ leads to improved quality with a less breathy voice. 
\begin{equation}
    \label{eq:refmse}
    \vspace{-0.15cm}
    \begin{alignedat}{3}
        L_{refmse} &= L_{refmse\_F0} + L_{refmse\_P} \\
        L_{refmse\_F0} &= \mathbb{E}_{(F0, \tilde{F0})}[\lambda_{F0} (F0 - \tilde{F0})^2] \\
        L_{refmse\_P} &= \mathbb{E}_{(P, \tilde{P})}\left[\frac{\lambda_{P}}{d_{P}} (P - \tilde{P})^2\right]
    \end{alignedat}
\end{equation}
where periodicity dimension $d_{P} = 12$ . We set $\lambda_{F0} = 50$ and $\lambda_{P} = 30$.

\textbf{Multi-window STFT loss (on vocoder output):} We calculate L1 loss between the amplified log magnitude STFT spectrograms of the reference audio $x$ and predicted audio $\tilde{x}$ as follows:
\begin{equation}
    \vspace{-0.15cm}
    \label{eq:mwstft}
        L_{mw\_stft}(G) = \mathbb{E}_{(x, \tilde{x})} \sum_{i=1}^{C}\frac{\lambda_{stft, i}||X_i - \tilde{X}_i||_1}{N_i} \\
\end{equation}
where $X_i = amp\_log(|STFT_i(x)|)$,  $\tilde{X}_i = amp\_log(|STFT_i(\tilde{x})|)$ , and $N_i$ is number of elements in the magnitude for the $i^{th}$ FFT size. $\lambda_{stft}s$ denote the loss weights for each of the $C=3$ FFT sizes of 512, 1024 and 2048, and are set at 25.7, 51.3 and 102.5 respectively. STFT extraction is done at a frame shift of 128 samples.  The $amp\_log$ operator amplifies the signal by 72dB, takes $log$ for the signal above $e$, and makes it linear below $e$. . This approach ensures that digital zero input maps to zero output, and we never get excessively large negative numbers after taking $log$. 
\begin{equation}
    \vspace{-0.15cm}
    amp\_log(y) = 
    \begin{cases}
     log(y * gain),  & \text{if } y * gain \geq e \\
     \frac{y * gain}{e} & \text{if } y * gain < e
    \end{cases}
\end{equation}

\textbf{Adversarial loss (on vocoder output):} The audio without adversarial loss sounds muffled due to over-smoothed spectral predictions in case of MSE-based losses. Analogous to the image domain \cite{isola2016},  adversarial loss helps to produce more realistic audio by making the vocal tract filter predictions sharper. The discriminators operate on the 257-dim magnitude spectrograms (512-point FFT) extracted at a frame shift of 128 samples. Since we do not model the phase from the data, the adversarial loss is on the magnitude spectrogram in contrast with other neural vocoders, such as MelGAN and HiFi-GAN, which apply it to the raw audio signal. 

Specifically, we have $K = 8$ discriminators, where each discriminator (except terminal ones) sees a 48-point band in the 257-dim spectrogram, with 8 overlapping points. The two terminal discriminators see a frequency band of 40 points because of overlap only on one side. Having multiple discriminators exploits the fact that the spectrogram has different characteristics in different frequency bands. We use the least squares adversarial losses \cite{mao2017} as defined below:
\begin{equation}
    \label{eq:advloss}
    \vspace{-0.15cm}
    \begin{alignedat}{3}
        L_{adv}(D_k) &= \mathbb{E}_{(X, \tilde{X})}\left[\frac{(D_k(X) - 1)^2}{N_k} + \frac{D_k(\tilde{X})^2}{N_k}\right] \\
        L_{adv}(G) &= \mathbb{E}_{\tilde{X}} \left[ \lambda_{adv}\sum_{k=1}^{K} \frac{(D_k(\tilde{X}) - 1)^2}{N_k}\right] \\
    \end{alignedat}
\end{equation}
where $D_k$ is the $k^{th}$ discriminator, $N_k$ is the number of elements in the $k^{th}$ STFT magnitude band the discriminator operates on. $X$ and $\tilde{X}$ are amplified log magnitude spectrograms of reference and predicted audio, as described for the multi-window STFT loss. We set the adversarial loss weight $\lambda_{adv} = 50$. All the  discriminators follow the same convolutional architecture; see Table \ref{tab:disc_arch}.  Each $D_k$ treats the input for its allocated frequency band as an image and classifies patches of size 5 x 31, equal to its receptive field. Like PatchGAN architectures \cite{isola2016}, we found that discriminators operating on smaller patches result in higher-quality audio than just one discriminator prediction for the entire input.

\par
Our final loss for training the generator is:
\begin{equation}
    \vspace{-0.15cm}
    \label{eq:finalloss}
    \begin{alignedat}{3}
        L(G) &= L_{refmse} + L_{mw\_stft} + L_{adv}(G) 
    \end{alignedat}
\end{equation}
\begin{table}[!]
 \vspace{-0.1in}
    \centering
    \caption{Discriminator architecture}
    \label{tab:disc_arch}
    \resizebox{0.47\textwidth}{!}{
        \begin{tabular}{c | c}
            \hline
            Layer & Details \\
            \hline
            Conv2d + LeakyReLU(0.2) & 3 x 3, stride = (1, 2),  \( \text{C}_{\text{out}} \) = 32 \\
            Conv2d + LeakyReLU(0.2) & 1 x 3, stride = (1, 2), \( \text{C}_{\text{out}} \)  = 64 \\
            Conv2d + LeakyReLU(0.2) & 1 x 3, stride = (1, 1), \( \text{C}_{\text{out}} \)  = 128 \\
            Conv2d + LeakyReLU(0.2)& 1 x 3, stride = (1, 1), \( \text{C}_{\text{out}} \)  = 256 \\
            Conv2d & 3 x 3, stride = (1, 1), \( \text{C}_{\text{out}} \)  = 1 \\
            \hline
        \end{tabular}
    }
    \vspace{-0.2in}
\end{table}

During DDSP vocoder training, reference values are used for the $F0$ feature to ensure that both predicted and reference audio waveforms are pitch-aligned for the STFT loss calculations, whereas for inference, the predicted $F0$ values are used for waveform generation. The initial learning rate of $G$ is set at $10^{-3}$ and all $Ds$ is set at $10^{-4}$, with both $G$ and $Ds$ using Adam optimizer \cite{kingma2017} with same parameters $\beta_1 = 0.9$ and $\beta_2 = 0.99$, and a weight\_decay of $10^{-6}$ and gradient clipping norm of 1.0. Convolution layers in the discriminators are regularized using weight normalization \cite{salimans2016}. Model training happens on a single Nvidia A100 GPU, with a batch size of 8 and a sequence length of 500 frames. We pretrain the generator using $L_{refmse}$ and $L_{mw\_stft}$ losses for an initial 5,000 iterations and then train the model for a total of 400,000 iterations. 
\section{Results}
\label{sec:exp}
\subsection{Experimental setup for comparison}
\label{ssec:expsetup}
For experiments, we use two internal studio-quality corpora: i) An American English female speaker with approximately 37 hours of audio and ii) An American English male speaker with approximately 12 hours of audio, with both corpora at 24KHz sampling rate. We hold out a test set of 42 utterances from each corpus for subjective evaluations.

We compare our DDSP system against five other TTS systems. Two are traditional DSP vocoder-based, while the other three systems are neural vocoders: MB-MelGAN, HiFi-GAN, and WaveRNN. The neural vocoders consume 13-dimensional Mel-Frequency Cepstral Coefficients (MFCC) as the spectral features, extracted with a 1024-point FFT and a frame shift of 128 samples. The DSP vocoder uses higher 80-dimensional $lmel_{psync}$ features because more spectral details are required to achieve a reasonable sounding synthesis. $lmel_{psync}$ features are extracted with a frame shift of one pitch period, a window size of two pitch periods, with an FFT size rounded to a power of two greater than or equal to window size for the voiced parts, and a 256-point FFT and a  frame shift of 128 samples for the unvoiced parts.

All vocoders also take $F0$ and the periodicity $P$ as inputs. These inputs are predicted from an acoustic model trained with only an L2 loss on the reference features, except for DSP Vocoder Adv, which also uses adversarial loss on the spectral feature for better perceptual quality. The acoustic model follows the architecture described in Table \ref{tab:am_arch}, with only the last linear layer modified according to the corresponding output dimension. We use the reference $F0$ and $dur$ features for all the TTS systems to remove the prosodic differences with the ground truth audio.

For MB-MelGAN, we use this open-source implementation \cite{pwgrepo}. The 128x upsampling is conducted through 4 upsampling layers with 4x, 2x, 2x, and 2x upsample factors, with output channels of the upsample networks as 256, 128, 64, and 32, respectively. The model outputs four sub-bands combined using PQMF synthesis filters. For WaveRNN, our implementation closely follows the original paper, while for HiFi-GAN, we based it off this repo \cite{hifiganrepo}. Both implementations have some architecture hyper-parameters differences compared to the original sources, so we compare with them only for audio quality.

\subsection{Speech Synthesis Quality Evaluation}
\label{sec:ssquality}
We conducted subjective MOS tests on the synthesized 42 test set utterances for each TTS system and corpus. Each TTS system received a total of 420 ratings, with each rating on a 5-point scale from 1 to 5. 

The evaluation results are shown in Table \ref{tab:mos}. DDSP Vocoder achieves a MOS score similar to other neural vocoders such as WaveRNN and HiFi-GAN and is rated higher than the performant on-device vocoder Multi-band MelGAN. DDSP Vocoder outperforms both the DSP vocoder baselines, with and without adversarial loss. Our approach shows that end-to-end optimization of the acoustic model with DSP vocoder helps to achieve natural-sounding audio\footnote{Audio Samples: \url{https://ddsp-vocoder.github.io/ddsp}}.

\begin{table}[!]
    \vspace{-0.2in}
    \centering
    \caption{Comparison of MOS scores with 95\% confidence intervals}
    \label{tab:mos}
    \resizebox{0.35\textwidth}{!}{
        \begin{tabular}{l | c | c}
            \hline
            Speaker & TTS Systems & MOS (CI) \\
            \hline
            \multirow{6}{*}{Female} & Ground Truth & $4.54 \pm 0.07$ \\
            & WaveRNN & $4.52 \pm 0.07$ \\
            & HiFi-GAN & $4.44 \pm 0.07$ \\
            & MB-MelGAN & $4.16 \pm 0.08$ \\
            & DSP Vocoder & $3.38 \pm 0.10$ \\
            & DSP Vocoder Adv & $3.69 \pm 0.09$ \\
            & \textbf{DDSP Vocoder} & $\boldsymbol{4.39 \pm 0.07}$ \\
            \hline
            \multirow{6}{*}{Male} & Ground Truth & $4.58 \pm 0.07$ \\
            & WaveRNN & $4.58 \pm 0.06$ \\
            & HiFi-GAN & $4.49 \pm 0.07$ \\
            & MB-MelGAN & $4.21 \pm 0.08$ \\
            & DSP Vocoder & $3.64 \pm 0.11$ \\
            & DSP Vocoder Adv & $3.83 \pm 0.10$ \\
            & \textbf{DDSP Vocoder} & $\boldsymbol{4.33 \pm 0.08}$ \\
            \hline
        \end{tabular}
    }
    \vspace{-0.1in}
\end{table}

\subsection{Model Complexity}
\label{sec:modelcomplexity}
\begin{table}[ht!]
    \vspace{-0.2in}
    \centering
    \caption{Performance comparison}
    \label{tab:complexity}
    \resizebox{0.47\textwidth}{!}{
        \begin{tabular}{l | c | c | c | c}
            \hline
            Model & GFLOPS &  RTF (vocoder) & RTF (overall) \\
            \hline
            MB-MelGAN & 5.2 & 0.102 & 0.138 \\
            DDSP Vocoder & 0.015 & 0.003 & 0.044 \\
            \hline
        \end{tabular}
    }
    \vspace{-0.1in}
\end{table}
We also evaluate the generation complexity and efficiency, summarized in Table \ref{tab:complexity}. All the RTF values are measured running single-threaded on an Intel(R) Xeon(R) CPU Gold 6138 @ 2GHz, with a C++ inference TTS pipeline, and all pytorch models are compiled to optimized torchscript \cite{jitwebpage}. The inference setup is the same till the acoustic model, with the vocoder being different across the two systems. For the FLOPS measurement for Multi-band MelGAN, we use ptflops tool \cite{ptflops} and modify it to output FLOPS instead of MACs (multi-add cumulation), accounting for 2 FLOPS per MAC. The FLOPS for the DDSP vocoder are manually estimated. The benchmark results show that the DDSP vocoder has 340 times lesser FLOPS, and 34 times lesser vocoder-only RTF with no parameters in the vocoder, than a production-grade neural vocoder system based on MB-MelGAN.

\section{Conclusion}
We present DDSP vocoder; a novel way of training of jointly optimizing an acoustic model and a DSP vocoder without using an engineered spectral feature, which leads to an audio quality close to high quality neural vocoders with much lower computation. In the future, we would like to extend the system to have multi-speaker and multi-lingual capabilities.

\vfill
\pagebreak

\bibliographystyle{IEEEbib}
\bibliography{references}

\end{document}